\documentclass{iaus}
\usepackage{graphicx}

\title[Theory 6.~~Anions \& Theory] 
{Theoretical Electronic and Rovibrational \\
Studies for Anions of Interest to the DIBs}

\author[Ryan C. Fortenberry]   
{Ryan C. Fortenberry
\thanks{Present address: Department of Chemistry, Georgia Southern University, Statesboro, GA 30460, U.S.A.}}

\affiliation{NASA Ames Research Center, Moffet Field, CA 94035, U.S.A.\\
email:{\tt ryan.c.fortenberry@nasa.gov}}

\pubyear{2013}
\volume{297}  
\pagerange{xxx--xxx}
\jname{The Diffuse Interstellar Bands}
\editors{J.~Cami \& N.~L.~J.~Cox, eds.}
\begin{document}

\maketitle

\begin{abstract}

The dipole-bound excited state of the methylene nitrile anion (CH$_2$CN$^-$)
has been suggested as a candidate carrier for a diffuse interstellar band (DIB)
at 803.8 nm.  Its corresponding radical has been detected in the interstellar
medium (ISM), making the existence for the anion possible.  This work applies
state-of-the-art $ab\ initio$ methods such as coupled cluster theory to
reproduce accurately the electronic excitations for CH$_2$CN$^-$ and the
similar methylene enolate anion, CH$_2$CHO$^-$.  This same approach has been
employed to indicate that 19 other anions may possess electronically excited
states, five of which are valence in nature.  Concurrently, in order to assist
in the detection of these anions in the ISM, work has also been directed
towards predicting vibrational frequencies and spectroscopic constants for
these anions through the use of quartic force fields (QFFs).  Theoretical
rovibrational work on anions has thus far included studies of CH$_2$CN$^-$,
C$_3$H$^-$, and is currently ongoing for similar systems.

\end{abstract}

\firstsection 
\section{Introduction}

Small molecular anions have been proposed as possible carriers of the diffuse
interstellar bands (DIBs) for over a decade now with the initial observation of
a coincidence between several bands of the open-shell C$_7^-$ radical anion
and at least five DIBs observed in \cite{Tulej98}.  Unfortunately, the spectral
features and, to a greater extent, the spectral shapes varied too greatly for
this anion to be considered a DIB carrier from the work by \cite{McCall01}.
Later, the dipole-bound excited state of CH$_2$CN$^-$ has been noted by
\cite{Sarre00} and, later, by \cite{Cordiner07} to possess an excitation energy
very close to a DIB at 8038 \AA.  This hypothesis has yet to be fully explored
and opens a new path by which DIB carriers may be explored.

A dipole-bound state of an anion is similar to a Rydberg state of a neutral.
However, a Rydberg state is bound by the interaction of the diffuse electron
with the now positively charged core of the molecule.  Hence, a
monopole-monopole interaction results where the core of the molecule functions
as a hydrogen nucleus.  Dipole-bound states of anions do not have the benefit
of such an interaction since removal of the electron from the anion results in
a neutral molecular core.  As such, most anionic electrons are weakly bound if
at all.  However, if the neutral core of the molecule, treated as an
independent system, has a large enough dipole moment, the electron may remain
bound within the system (see \cite{Simons08}).  The minimum dipole moment
necessary for a neutral molecule to bind an extra electron in a $\sigma$-type
orbital is 1.625 D as originally derived by \cite{Fermi47}.  Physical
conditions actually place this threshold at more than 2.5 D as detailed by
\cite{Gutsev95}.  Since these dipole-bound states, like Rydberg states, exist
on the cusp of the dissociation continuum, only one such state can be present
for a given anion unless the dipole moment is closer to 10 D.  As a result, the
electronic spectra of anions are much simpler than cations or neutral
molecules.

CH$_2$CN$^-$ is a closed-shell, valence anion in its $\tilde{X}\ ^1A'$ state.
As a result, electronic excitation is possible into a dipole-bound excited
state and is known to exist experimentally at 1.543 eV (803.8 nm) from
\cite{Lykke87}.  CH$_2$CHO$^-$ is also a known closed-shell valence ground
state anion and to possess a dipole-bound excited state at 1.759 eV (704.9 nm)
from \cite{Mullin92}.  Additionally, the CH$_2$CN radical has been observed in
the interstellar medium (ISM), making the existence of CH$_2$CN$^-$ well within
reason.  With these anions as benchmarks, high-level, $ab\ initio$, quantum
chemical computations are employed to explore the electronic spectra of several
small anions.

\section{Computational Methods}

As outlined previously by Fortenberry \& Crawford (2011a,b) and
\cite{Fortenberry133dbs}, energy differences from geometry optimizations of the
ground states for the anion and neutral radical with CCSD(T)/aug-cc-pVTZ give
adiabatic electron binding energies (eBEs), which is the energy necessary to
remove the electron from the system.  EOM-CCSD/d-aug-cc-pVDZ geometery
optimizations of the ground and first excited states of the anion produce
adiabatic excitation energies from the energy differences of these optimized
geometries.  Even though this is a best-to-best comparison that should prove
more accurate, geometry optimization is the manipulation of several variables.
As a result, vertical computations are also employed which freeze the geometry
at a select point (in this case the CCSD(T)/aug-cc-pVTZ geometry for the ground
state of the anion) and manipulate the wavefunction either for electron removal
or excitation.  The vertical excitation energies are computed with EOM-CCSD
with an increasing level of diffuseness in the spatial extent of the cc-pVDZ
and cc-pVTZ basis sets and are derived in an even-tempered fashion.
Convergence to an excitation energy only until highly diffuse basis sets are
used implies that the electron probability density is further away from the
core of the molecule.  This indicates, for the present purposes, the presence
of a dipole-bound excited state.

Additionally, highly-accurate rovibrational computations are employed on a
selection of anions that possess electronically excited states.  The
spectroscopic constants and vibrational frequencies provide more data to assist
in their interstellar detection.  This procedure involving quartic force fields
(QFFs), which is a numerical solving of the nuclear potential energy function
as a fourth-order Taylor series expansion, is detailed elsewhere:
\cite{Huang08} and \cite{Fortenberry11HOCO}.  Accuracies for computed
spectroscopic data compared with experiment for some closed-shell systems are
known to be better than 0.1\% as is the case with C$_3$H$_3$$^+$ reported by
\cite{Huang11}; HOCO$^+$ from \cite{Fortenberry12hococat}; and NNOH$^+$ given
in \cite{Huang13NNOH+}.

\section{Discussion}

As shown in Table \ref{ebe}, the adiabatic eBE computations place the
CH$_2$CN$^-$ at 1.48 eV, 0.06 eV less than the experimental value, while the
adiabatic CH$_2$CHO$^-$ eBE is computed to be 1.77 eV or 0.05 eV less than
experiment.  The adiabatic first excited state energy for CH$_2$CN$^-$ is 1.49
eV, 0.04 eV less than experiment.  The same value for CH$_2$CHO$^-$ is 1.77 eV
or 0.01 eV more than experiment.  Electronic transitions within 0.1 eV of
experiment are the best that can be expected even for the EOM-CCSD method, and
these computations for the benchmarking anions highlight the level of accuracy
expected.

\begingroup
\begin{table}[h]

\caption{Computed dipole moments (in Debye for the corresponding neutral
radical), eBEs (in eV), and first adiabatic excited state transition energies
(eV) and wavelengths (nm) for several anions.}

\centering
\begin{tabular}{l c c c c c}
\hline

Molecule \hspace{0.01 in} & Dipole & eBE & Transition & Energy & $\lambda$ \\

\hline

CH$_2$CN$^-$ & 3.509 & 1.48 & 1$\ ^1B_1 \leftarrow 1\ ^1A'$ & 1.49 & 830 \\

CH$_2$SiN$^-$ & 4.110 & 2.49 & 1$\ ^1B_1 \leftarrow 1\ ^1A_1$ & 2.10 & 590
\\

SiH$_2$CN$^-$ & 3.524 & 2.31 & 2$\ ^1A'\leftarrow 1\ ^1A'$ & 2.39 & 519 \\

CH$_2$CHO$^-$ & 2.921 & 1.77 & 1$\ ^1A''\leftarrow 1\ ^1A'$ & 1.77 & 702 \\

CH$_2$SiHO$^-$ & 4.452 & 2.45 & 1$\ ^1A''\leftarrow 1\ ^1A'$ & 2.47 & 503
\\

SiH$_2$CHO$^-$ & 2.391 & 1.93 & 2$\ ^1A\leftarrow 1\ ^1A$ & 2.20 & 564 \\

CH$_2$ON$^-$ & 2.335 & 0.46$^a$ & $2\ ^1A \leftarrow 1\ ^1A$ & 0.89 & 1396 \\

CH$_2$SN$^-$ & 2.703 & 1.98 & $2\ ^1A' \leftarrow 1\ ^1A'$ & 1.95 & 634 \\

CH$_2$NO$^-$ & 2.317 & 1.42 & $2\ ^1A' \leftarrow 1\ ^1A'$ & 1.55 & 799
\\

CH$_2$PO$^-$ & 2.477 & 2.82 & $1\ ^1A'' \leftarrow 1\ ^1A'$ & 2.89 & 429 \\

C$_3$N$^-$ & 2.889 & 4.38 & 2$\ ^1\Sigma^+ \leftarrow 1\ ^1\Sigma^+$ & 4.71 &
263 \\

SiCCN$^-$ & 4.010 & 3.44 & 2$\ ^1A' \leftarrow 1\ ^1\Sigma^+$ & 3.13 & 397 \\

CCSiN$^-$ & 0.672$^b$ & 3.93 & 1$\ ^1A'' \leftarrow 1\ ^1\Sigma^+$ &  2.49 &
498\\

SiCN$^-$ & 3.188 & 1.27 & 1$\ ^1\Pi \leftarrow 1\ ^1\Sigma^+$ & 1.37 & 906 \\

SiNC$^-$ & 2.709 & 0.92 & 1$\ ^1\Pi \leftarrow 1\ ^1\Sigma^+$ & 1.07 & 1158 \\

CN$^-$ & 1.471 & 3.82 & 1$\ ^1\Pi \leftarrow 1\ ^1\Sigma^+$ & 4.43 & 280 \\

SiN$^-$ & 2.585 & 2.97 & 2$\ ^1\Sigma^+ \leftarrow 1\ ^1\Sigma^+$ & 3.23 &
383 \\

C$_2$F$^-$ & 1.075 & 3.17 & 2$\ ^1A' \leftarrow 1\ ^1\Sigma^+$ & 3.51 & 354
\\

HCCO$^-$ & 2.170 & 2.39 & 1$\ ^1A''\leftarrow 1\ ^1A'$ & 2.37 & 523 \\

CCOH$^-$ & 4.401 & 2.52 & 1$\ ^1A''\leftarrow 1\ ^1A'$ & 2.43 & 511 \\

CCSH$^-$ & 4.492 & 2.86 & $1\ ^1A'' \leftarrow 1\ ^1A'$ & 2.82$^c$ & 440 \\

C$_3$H$^-$ & 3.409 & 1.83 & $1\ ^1A'' \leftarrow 1\ ^1A'$ & 0.93 & 1328 \\

CCSiH$^-$ & $\sim$3$^c$ & 3.11 & $1\ ^1A'' \leftarrow 1\ ^1A'$ & 2.06 & 602
\\

CCNH$_2^-$ & 5.903 & 1.97 & $1\ ^1B_1 \leftarrow 1\ ^1A'$ & 2.00 & 620 \\

CCPH$_2^-$ & 4.759 & 3.21 & $2\ ^1A' \leftarrow 1\ ^1A'$ & 3.16 & 392 \\

BH$_3$NH$_2^-$ & 2.524 & 2.28 & $2\ ^1A' \leftarrow 1\ ^1A'$ & 2.53 & 490 \\

BH$_3$PH$_2^-$ & 2.889 & 2.74 & $2\ ^1A' \leftarrow 1\ ^1A'$ & 2.78 & 446 \\

AlH$_3$NH$_2^-$ & 2.579 & 3.22 & $2\ ^1A' \leftarrow 1\ ^1A'$ & 3.17 & 391
\\

AlH$_3$PH$_2^-$ & 2.167 & 3.25 & $2\ ^1A' \leftarrow 1\ ^1A'$ & 3.21 & 387
\\

\hline

\end{tabular}
\label{ebe}

{\raggedright

$^a$This value is from the non-minimum geometry but functions as an upper bound
estimate.\\

$^b$The lowest energy isomer of the radical is a cyclic ring with the N bonded
to the Si.  The linear isomer has a larger dipole moment, but its stability is
uncertain.\\

$^c$These values are upper-(CCSH$^-$) or lower-(CCSiH$^-$)bound estimates.\\}

\end{table}
\endgroup

Theoretical and computational chemistry are uniquely suited to explore
properties of various molecules where laboratory examination can prove quite
difficult.  Hence, knowing the accuracies of this approach, new anions can be
explored for their excited state properties.  The anions must be closed-shell,
valence anions in order to possess a higher electronic state.  The
corresponding neutral radicals must have large dipole moments of greater than
at least 2 D.  Small anions are preferrable for the methods required and for
the astrophysical conditions expected.  The list of anions explored to date
along with their computed properties is given in Table \ref{ebe} and is a
compilation of data from the studies undertaken by Fortenberry \& Crawford
(2011a,b) and \cite{Fortenberry133dbs}.  Of those anions listed, 19 new ones
clearly have adiabatic excitation energies that are less than or no more than
0.1 eV above the eBEs.  These include CH$_2$SiN$^-$, SiH$_2$CN$^-$,
CH$_2$SiHO$^-$, SiN$^-$, CCOH$^-$, HCCO$^-$, SiCCN$^-$, CSiCN$^-$, CCSiN$^-$,
SiNC$^-$, CH$_2$SN$^-$, C$_3$H$^-$, CCSiH$^-$, CCSH$^-$, CCNH$_2^-$,
CCPH$_2^-$, BH$_3$PH$_2^-$, AlH$_3$NH$_2^-$, and AlH$_3$PH$_2^-$.  Most of
these electronic transitions are within the wavelength range ($\sim 1 - 3.5$
eV) of the DIBs and represent a new set of molecules whose relationship to the
DIBs and the chemistry of the ISM have not been explored.

\begin{table}

\caption{EOM-CCSD vertical excitation energies (in eV), oscillator
strengths,$^a$ and vertical electron binding energies (in eV) from
ground state CCSD(T)/aug-cc-pVTZ geometries for several basis
sets.$^b$}

\begin{tabular}{l l | c c c c | c c c | c || c}
\hline
\label{basis}

Molecule & Transition & pVDZ & apVDZ & dapVDZ & tapVDZ & pVTZ & apVTZ
& dapVTZ & $f$ & eBE$^c$ \\

\hline

CH$_2$SiN$^-$
 & 1$\ ^1B_1\leftarrow 1\ ^1A_1$ & 2.90 & 2.24 & 2.12 & 2.12 & 2.69 & 2.29 & 2.23 & 3$\times10^{-3}$ & 2.34 \\

 & 2$\ ^1B_1\leftarrow 1\ ^1A_1$ & 6.44 & 3.38 & 2.59 & 2.39 & 5.87 & 3.33 & 2.73 & 4$\times10^{-3}$ & \\

SiCCN$^-$
 & 2$\ ^1\Sigma^+ \leftarrow 1\ ^1\Sigma^+$ & 3.58 & 3.30 & 3.27 & 3.27 & 3.41 & 3.24 & 3.23 & 1$\times10^{-5}$ & 3.33 \\

 & 1$\ ^1\Pi \leftarrow 1\ ^1\Sigma^+$ & 7.17 & 4.50 & 3.50 & 3.36 & 6.45 & 4.36 & 3.62 & 1$\times10^{-2}$ & \\

CCSiN$^-$
 & 1$\ ^1\Pi \leftarrow 1\ ^1\Sigma^+$ & 4.15 & 3.75 & 3.71 & 3.71 & 3.89 &
3.71 & 3.70 & 5$\times10^{-2}$ & 4.51 \\

 & 2$\ ^1\Sigma^+ \leftarrow 1\ ^1\Sigma^+$ & 4.35 & 4.22 & 4.21 & 4.21 & 4.23
& 4.17 & 4.16 & 1$\times10^{-7}$ & \\

 & 2$\ ^1\Pi \leftarrow 1\ ^1\Sigma^+$ & 6.20 & 5.71 & 4.82 & 4.60 & 5.99 &
5.68 & 4.93 & 2$\times10^{-5}$ & \\

C$_3$H$^-$
 & $1\ ^1A'' \leftarrow 1\ ^1A'$ & 1.21 & 1.15 & 1.14 & 1.14 & 1.17 & 1.14 &
1.13 & 2$\times 10^{-3}$ & 2.34 \\

 & $2\ ^1A' \leftarrow 1\ ^1A'$ & 6.38 & 3.01 & 2.46 & 2.37 & 5.68 &
3.03 & 2.61 & 3$\times 10^{-3}$ & \\

CCSiH$^-$
 & $1\ ^1A'' \leftarrow 1\ ^1A'$ & 2.09 & 2.05 & 2.05 & 2.05 & 2.11 & 2.07 &
2.07 & 1$\times 10^{-2}$ & 3.27\\

 & $2\ ^1A' \leftarrow 1\ ^1A'$ & 6.40 &
4.02 & 3.32 & 3.26 & 6.00 & 3.92 & 3.45 & 2$\times 10^{-2}$ & \\

\hline
\end{tabular}

$^a$Oscillator strengths ($f$ values) reported are for CCSD/d-aug-cc-pVTZ.\\
$^b$Dunning's correlation consistent basis sets are abbreviated, {\em e.g.}
t-aug-cc-pVDZ is tapVDZ.\\
$^c$Computed with EOMIP-CCSD/t-aug-cc-pVDZ.\\

\end{table}

Closer examination of the difference in the excitation energies and the eBEs
discloses that five of these anions have a substantial difference between the
two values.  Since dipole-bound states are Rydberg-like, the eBE and the
dipole-bound excited state energy must be nearly coincident.  This is observed
for the two benchmarking anions.  However, CH$_2$SiN$^-$, SiCCN$^-$, CCSiN$^-$,
C$_3$H$^-$, and CCSiH$^-$ have differences of several tenths of an eV.
Vertical excitation energies are useful in classifying the nature of excited
states since only the wavefunctions are manipulated.  These are given in Table
\ref{basis} for both the cc-pVDZ and cc-pVTZ families of increasingly diffuse
basis sets.  From this table it is clear that the first excited state for each
of these anions is not dipole-bound since the majority of the spatial extent of
the required basis set is described by the aug-cc-pVXZ basis.  The large
decline in the excitation energy as the basis sets become more diffuse for the
higher-energy excitation clearly indicates the dipole-bound state.
Additionally, these energies converge to within 0.1 eV of the EOMIP vertical
eBE.  Furhermore, CCSiN$^-$ has two valence excited states below the eBE, the
first small anion studied where such a result has been given.  All of these
anions with valence excitations contain silicon except C$_3$H$^-$.  Although it
is currently unclear as to why these anions possess valence excited states and
others do not, especially for most of their first-row analogues, this
phenomenon appears to be a combination of special $\pi \rightarrow \pi$ (not
$\pi \rightarrow \pi^*$) excitations involving partially filled $\pi$-type
orbitals as well as the charge separation brought about by the larger Si atom
and its propensity only to form $\sigma$ bonds.

Multiple excited states of anions further gives new areas for exploration of
potential DIB carriers.  Conclusive assignments of excitation energies for any
anions will probably result from laboratory experiments, but these theoretical
computations have highlighted new physics that have been largely unexplored.
Additionally, anions are attractive DIB carriers since most have only one
excited state, and it is believed that most of the DIBs are not correlated.
The excitation energies for most anions (especially those containing only the
more astronomically abundant H, C, O, and N) are more toward the red end of the
spectrum.  The valence state of C$_3$H$^-$ should be found at more than 13000
\AA, for example.  This reddening is consistent with a majority of the known
DIBs and is further supported by new DIBs discovered by \cite{Geballe11} beyond
10000 \AA.

Excitation energies are useful, but detection of these anions in the ISM may
first be necessary with vibrational or rotational spectroscopy.  As such, the
fundamental vibrational frequencies and spectroscopic constants of the DIB
candidate CH$_2$CN$^-$ are given in \cite{Fortenberry13CH2CN-}.  More
excitingly, since the attribution of C$_3$H$^+$ lines observed in the
Horsehead nebula photodissociation region (PDR) has been questioned by
\cite{Huang13C3H+}, $1\ ^1A'$ C$_3$H$^-$ is the more likely candidate for those
rotational lines from a spectroscopic perspective as suggested by
\cite{Fortenberry13C3H-}.  With the existence of the valence state for this
anion, it is likely that electron attachment rates in this region could be
greater than the dissociation rate such that a steady state of C$_3$H$^-$ could
be present.  This would be the first small anion known to possess a valence
excited state to be detected in the ISM.


\section{Acknowledgements}

Travel and research funds were provided to RCF by the NASA Postdoctoral Program
administered through Oak Ridge Associated Universities.  Prof.~T.~Daniel
Crawford of Virginia Tech is acknowledged for his assistance with the excited
state studies as well as for the use of the computing resources.  Additionally,
Drs.~Xinchuan Huang and Timothy J.~Lee of the SETI Institute and NASA Ames
Research Center, respectively, were instrumental in the rovibrational studies
undertaken.

\end{document}